\documentclass[reprint,aps,prb]{revtex4-2}

\usepackage{xcolor}
\usepackage{graphicx}
\usepackage{color}
\usepackage{amsthm, amssymb, amsmath}
\usepackage{amsfonts}
\usepackage{mathtools}
\usepackage{bm}
\usepackage{physics}
\usepackage{ulem}
\usepackage{ascmac}
\usepackage{comment}
\usepackage{empheq}
\usepackage[thicklines]{cancel}
\usepackage{enumerate}
\usepackage{cases}
\usepackage{mathrsfs}
\usepackage{booktabs}
\usepackage{multirow}
\usepackage[version=4]{mhchem}
\usepackage[caption=false, position=top]{subfig}
\allowdisplaybreaks

\newcommand{\Def}{\coloneq}

\newcommand{\up}{\uparrow}
\newcommand{\down}{\downarrow}

\newcommand{\Z}{\mathbb{Z}}

\newcommand{\kx}{k_\mathrm{x}}
\newcommand{\ky}{k_\mathrm{y}}
\newcommand{\kz}{k_\mathrm{z}}

\newcommand{\sx}{s_\mathrm{x}}
\newcommand{\sy}{s_\mathrm{y}}
\newcommand{\sz}{s_\mathrm{z}}

\newcommand{\rhox}{\rho_\mathrm{x}}
\newcommand{\rhoy}{\rho_\mathrm{y}}
\newcommand{\rhoz}{\rho_\mathrm{z}}

\newcommand{\sigmax}{\sigma_\mathrm{x}}
\newcommand{\sigmay}{\sigma_\mathrm{y}}
\newcommand{\sigmaz}{\sigma_\mathrm{z}}

\newcommand{\taux}{\tau_\mathrm{x}}
\newcommand{\tauy}{\tau_\mathrm{y}}
\newcommand{\tauz}{\tau_\mathrm{z}}

\DeclareMathOperator{\sgn}{sgn}

\newcommand{\M}{\bar{\mathrm{M}}}
\newcommand{\X}{\bar{\mathrm{X}}}

\renewcommand{\wp}{\mathrm{wp}}
\newcommand{\scwp}{\mathrm{scwp}}

\newcommand{\occ}{\mathrm{occ}}

\usepackage{hyperref}
\hypersetup{
    citecolor = blue,
    colorlinks = true,
    urlcolor = blue,
    linkcolor = blue
}

\usepackage{orcidlink}

\begin{document}

\title{
    Double-twisted surface spectrum from hybridized Majorana Kramers pairs and wallpaper fermions
}
\author{Kaito Yoda \orcidlink{0009-0006-5607-6364}}
\email{yoda.kaito.c1@s.mail.nagoya-u.ac.jp}
\affiliation{Department of Physics, Nagoya University, Nagoya 464-8602, Japan}
\author{Ai Yamakage \orcidlink{0000-0003-4052-774X}}
\affiliation{Department of Physics, Nagoya University, Nagoya 464-8602, Japan}

\date{\today}

\begin{abstract}
    We theoretically investigate the superconducting surface states of topological nonsymmorphic crystalline insulators with wallpaper fermions, which are surface quasiparticles protected by a wallpaper group $p4g$ symmetry, based on a tight-binding model for the space group $P4/mbm$ (No.~127).
    A symmetry-based analysis shows that four types of on-site pair potentials are allowed.
    Using the symmetries of the wallpaper group $p4g$ and the one-dimensional topological invariants, we clarify that for the $\mathrm{A_{1u}}$ representation, wallpaper fermions and two Majorana Kramers pairs coexist, and hybridization between them gives rise to a \textit{double-twisted} surface state and produces four peaks in the surface density of states.
    We further find that the mirror Chern number vanishes, indicating that our system realizes mirror-helicity-free surface states.
    This distinguishes superconducting wallpaper fermions from the other superconducting topological (crystalline) insulators, such as $\mathrm{Cu}_x\mathrm{Bi}_2\mathrm{Se}_3$ and $\mathrm{Sn}_{1-x}\mathrm{In}_x\mathrm{Te}$.
\end{abstract}

\maketitle

\section{Introduction} \label{sec:introduction}

Unconventional superconductors can host gapless Andreev bound states (ABSs) at their boundaries, including chiral, helical, and flat-band edge modes~\cite{qi2011topological, mizushima2016symmetry, sato2016majorana, chiu2016classification, sato-ando2017topological, aguado2017majorana}.
These boundary excitations originate from nontrivial bulk topology of the Bogoliubov--de Gennes (BdG) Hamiltonian and therefore constitute a hallmark of topological superconductivity.
Among these gapless boundary states, the self-conjugate ones are regarded as Majorana fermions that obey non-Abelian statistics~\cite{ivanov2001non-abelian}, offering a route toward topological quantum computation~\cite{kitaev2006anyons, nayak2008non-abelian, alicea2012new, beenakker2013search}.

The surface physics becomes richer when superconductivity is realized in topological materials, such as topological insulators~\cite{yonezawa2018nematic}, topological crystalline insulators~\cite{sasaki2012odd, novak2013unusual, he2013full, hashimoto2015surface, kawakami2018topological}, and topological semimetals~\cite{kobayashi2015topological, lu2015crossed, aggarwal2016unconventional, wang2016observation, wang2017discovery}.
For superconducting topological insulators such as $\mathrm{Cu}_x\mathrm{Bi}_2\mathrm{Se}_3$~\cite{yonezawa2018nematic}, mirror-parity-odd pair potentials cannot open a superconducting gap for surface Dirac fermions in the normal state.
As a result, in the superconducting states, the Dirac fermions can hybridize with Majorana Kramers pairs, which leads to an anomalous \textit{twisted} surface dispersion with multiple zero-energy crossings distinct from the purely linear spectrum of an isolated Majorana cone~\cite{hao2011surface, hsieh2012Majorana, yamakage2012theory, yamakage2013theory}.
A related mechanism has been discussed for superconducting topological crystalline insulators such as $\mathrm{Sn}_{1-x}\mathrm{In}_x\mathrm{Te}$~\cite{hashimoto2015surface}, where multiple normal-state surface Dirac cones further enrich the superconducting surface spectra.
These examples illustrate a general mechanism: gapless surface bands derived from the normal state can hybridize with superconducting ABSs and generate unconventional surface dispersions, depending on the type of topological surface states in the normal state.

Wallpaper fermions provide a natural unexplored platform in which to test this physics~\cite{wieder2018wallpaper,ryu2020wallpaper,zhou2021glide,mao2022third,hwang2023magnetic,mizuno2023hall,mizuno2025magnon,yoda2026superconducting}.
They are surface states of topological nonsymmorphic crystalline insulators protected by time-reversal symmetry and two orthogonal glide symmetries.
Unlike conventional Dirac surface states, wallpaper fermions exhibit fourfold degeneracies enforced by nonsymmorphic symmetry~\cite{wieder2018wallpaper}.
This feature suggests that hybridized states between wallpaper fermions and superconducting ABSs can give rise to novel quasiparticle states.

In this paper, we theoretically investigate surface states of wallpaper fermion systems with bulk superconductivity within a tight-binding model for on-site pair potentials.
We first classify the four on-site pairing channels permitted by the crystalline symmetries and identify the $\mathrm{A_{1u}}$ channel as the one in which gapless wallpaper fermions coexist with two Majorana Kramers pairs.
For the $\mathrm{A_{1u}}$ pairing channel, we compute the surface spectral function by using the recursive Green's function method~\cite{umerski1997closed}.
We find that the hybridization between wallpaper fermions and Majorana Kramers pairs gives rise to \textit{double-twisted} surface spectra and generates four peaks in the surface density of states (DOS).
We further show that these surface spectra are characterized by the absence of mirror helicity that distinguishes them from those of superconducting topological (crystalline) insulators~\cite{hsieh2012Majorana, hashimoto2015surface}.

The paper is organized as follows.
In Sec.~\ref{sec:model}, we construct a tight-binding model for the topological nonsymmorphic crystalline insulators with wallpaper fermions and classify the on-site pair potentials allowed by the crystalline symmetries.
In Sec.~\ref{sec:surface_state}, we analyze the hybridized states for the pairing symmetries in which wallpaper fermions and Majorana Kramers pairs coexist, and discuss the surface DOS.
In Sec.~\ref{sec:discussion}, we compare our results with previous theoretical studies of superconducting topological insulators~\cite{hsieh2012Majorana, hashimoto2015surface} by using the mirror Chern number.
Finally, Sec.~\ref{sec:conclusion} summarizes this paper.

\section{Model} \label{sec:model}
Wallpaper fermions are symmetry-protected surface states that arise in nonsymmorphic crystalline systems with two orthogonal glide symmetries, as realized in the wallpaper group $pgg$ or $p4g$~\cite{wieder2018wallpaper}.
Due to these nonsymmorphic symmetries, wallpaper fermions exhibit a fourfold degeneracy at the $\M$ point and a twofold degeneracy along the $\X\M$ lines.
In this work, we focus on the $p4g$ case, which possesses higher symmetry than $pgg$.
Wallpaper group $p4g$ contains fourfold and twofold rotation symmetries $\{4^{\pm}|0\ 0\}$ and $\{2|0 \ 0\}$ in addition to two orthogonal glide symmetries $\{m_{10}|1/2\ 1/2\}$ and $\{m_{01}|1/2\ 1/2\}$.
These symmetries also imply the presence of mirror symmetries $\{m_{11}|1/2\ 1/2\}$ and $\{m_{1\bar{1}}|1/2\ 1/2\}$.
In this section, we construct the model of wallpaper fermions for normal and superconducting states.

\subsection{Tight-binding Hamiltonian for wallpaper fermions} \label{sec:TB-wp}

\begin{figure}
    \centering

    \subfloat[]{
        \includegraphics[keepaspectratio,scale=0.65]{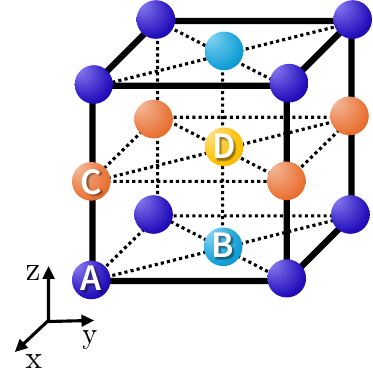}
        \label{fig:model}
    }\hfil
    \subfloat[]{
        \includegraphics[keepaspectratio,scale=0.35]{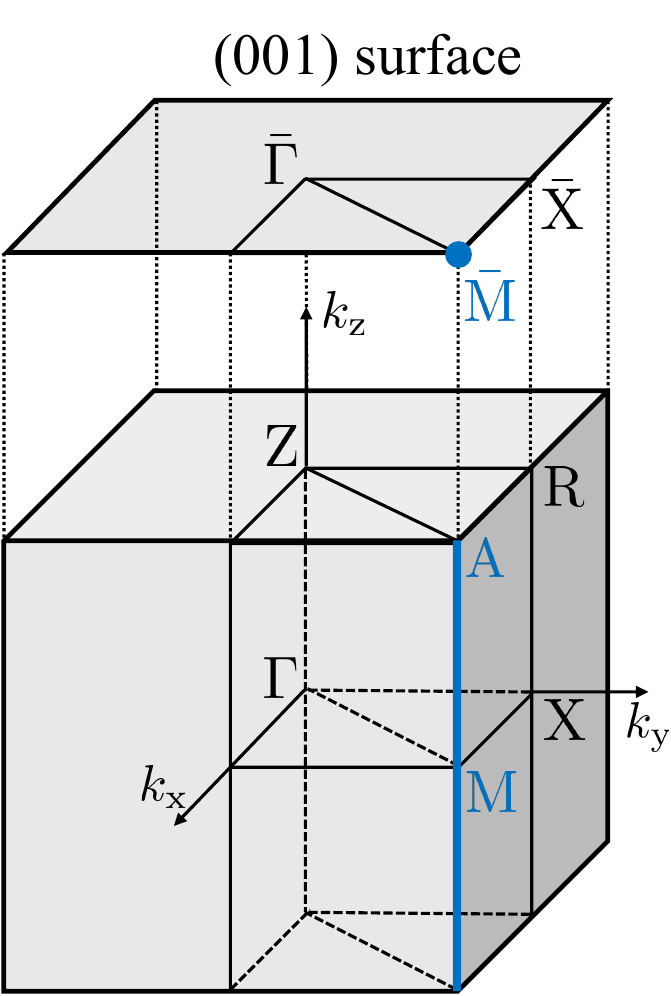}
        \label{fig:BZ}
    }

    \caption{(Color online)
        (a) Tetragonal lattice model for wallpaper fermions~\cite{wieder2018wallpaper}.
        (b) Brillouin zone for space group $P4/mbm$ and $(001)$ surface Brillouin zone~\cite{Aroyo2011-cr, Aroyo2006-bi1, Aroyo2006-bi2, Aroyo2014brillouin}.
    }
    \label{fig:model_BZ}
\end{figure}
We begin by introducing a tight-binding model for topological nonsymmorphic crystalline insulators with wallpaper fermions.
We employ the tetragonal lattice model shown in Fig.~\ref{fig:model_BZ}\subref{fig:model}, originally proposed in Ref.~\onlinecite{wieder2018wallpaper}.
The crystal structure consists of two types of layers stacked along the $z$ axis: one layer contains the A and B sublattices, while the other contains the C and D sublattices.
The bulk crystal symmetry is described by the space group $P4/mbm$ (No.~127).
The $(001)$ surface preserves the symmetries of the wallpaper group $p4g$.
The bulk and surface Brillouin zones are shown in Fig.~\ref{fig:model_BZ}\subref{fig:BZ}.
This model has three internal degrees of freedom: spin ($\up$ and $\down$), layer sublattice [A(B) and C(D)], and in-plane sublattice [A(C) and B(D)].
These are represented by Pauli matrices $s_\nu,\ \rho_\nu,$ and $\sigma_\nu$ ($\nu=0,\mathrm{x,y,z}$), respectively.
Hereafter, we set the lattice constants to unity.

\begin{figure}
    \centering
    \includegraphics[keepaspectratio,scale=0.33]{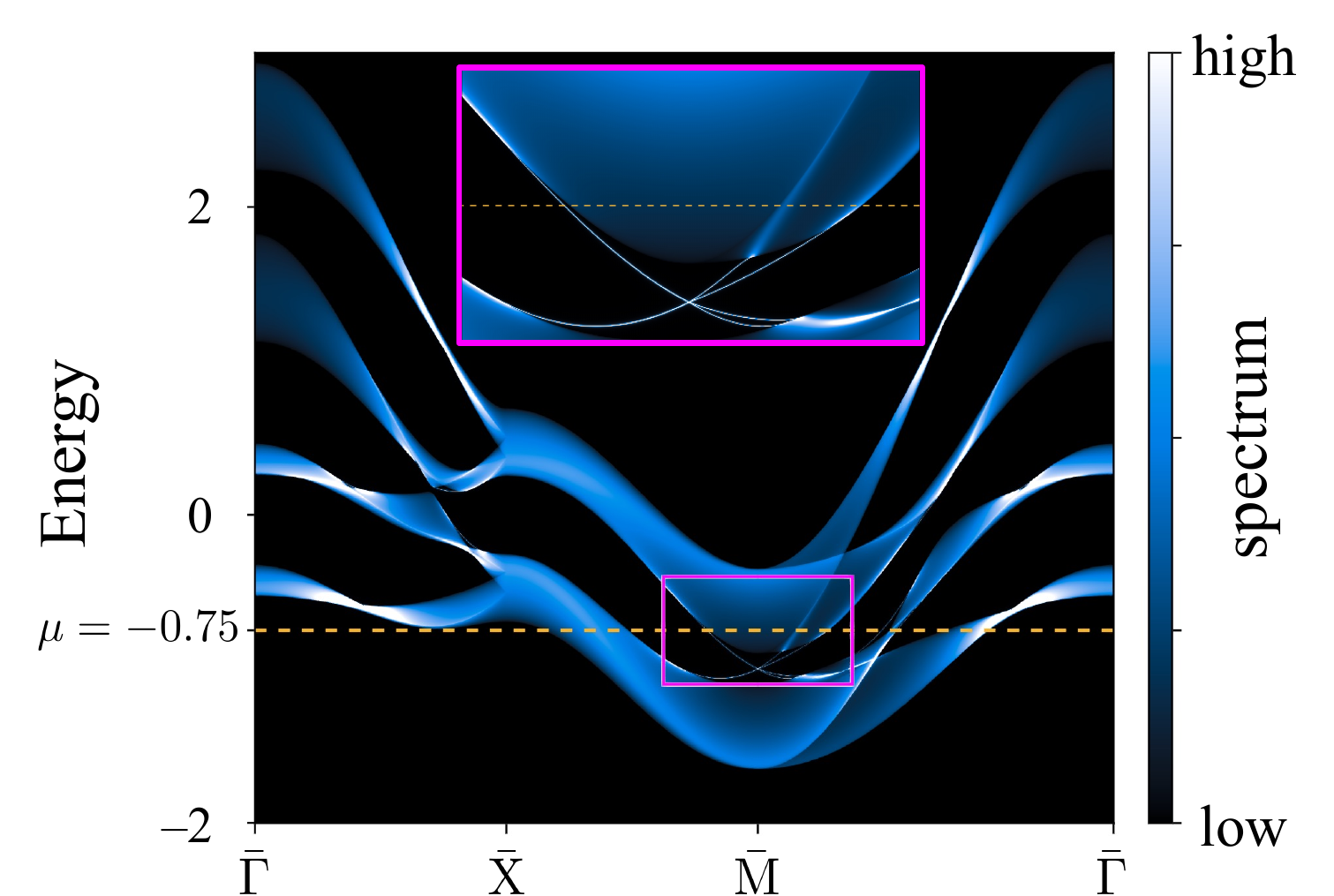}
    \caption{(Color online)
        Surface energy spectrum for a semi-infinite system with the $(001)$ surface on the C--D layers.
        The inset shows an enlarged view near the $\M$ point.
        A fourfold degeneracy at the $\M$ point and a twofold degeneracy along the $\X\M$ line are confirmed.
        The parameters are fixed as $t_1=1,\ v_{\mathrm{r}1}=0.25,\ v_{\mathrm{s}1}=0.2,\ t_2=0.5,\ v_{\mathrm{s}2}=-0.2,\ v'_{\mathrm{s}2}=0.15,\ u_1=0.3,\ u_2=-0.45,\ w_1=0.2,\ w_2=0.05,\ w_3=0.05,\ w_4=-0.02,\ w_5=0.11,\ w_6=-0.08,\ v_1=0.2,\ v_2=0.1,\ p_1=0,\ p_2=0.3$.
        The dashed orange line marks the chemical potential $\mu=-0.75$ used in Sec.~\ref{sec:hybridization}.
    }
    \label{fig:wp_ene}
\end{figure}
We now construct a tight-binding Hamiltonian $H_\wp(\bm{k})$ consistent with the space group $P4/mbm$.
For each symmetry operation $g$ in the space group $P4/mbm$, the Hamiltonian satisfies $D_{\bm{k}}(g)H_\wp(\bm{k})D_{\bm{k}}(g)^\dag=H_\wp(g\bm{k})$, where $D_{\bm{k}}(g)$ is the representation matrix of $g$ at momentum $\bm{k}$.
The representations of generators for this lattice system are derived as follows (see Appendix~\ref{app:rep})~\cite{satow2025symmetry}:
\begin{align}
     & D_{\bm{k}}(\{4^+_{001}|\bm{0}\})=\frac{1}{\sqrt{2}}(s_0-i\sz)\rho_0\sigma_0, \label{eq:four-rot} \\
     & D_{\bm{k}}(\{2_{100}|1/2\ 1/2\ 0\})=-ie^{-i\frac{\kx-\ky}{2}}\sx\rhox\sigmax, \label{eq:x-screw} \\
     & D_{\bm{k}}(\{I|\bm{0}\})=s_0\rhox\sigma_0. \label{eq:inversion}
\end{align}
The time-reversal symmetry imposes $\Theta H_\wp(\bm{k})\Theta^{-1}=H_\wp(-\bm{k})$, with $\Theta=-i\sy\mathcal{K}$ and $\mathcal{K}$ denotes complex conjugation.
For in-plane directions, we include the nearest-neighbor hopping
\begin{align}
    H^{(1)}_{\parallel}(\bm{k})\Def
     & t_1\cos{\frac{\kx}{2}}\cos{\frac{\ky}{2}}[s_0\rho_0\sigmax] \notag              \\
     & +v_{\mathrm{r}1}\cos{\frac{\kx}{2}}\cos{\frac{\ky}{2}}[\sz\rho_0\sigmay] \notag \\
     & +v_{\mathrm{s}1}\sin{\frac{\kx}{2}}\cos{\frac{\ky}{2}}[\sy\rhoz\sigmax] \notag  \\
     & -v_{\mathrm{s}1}\cos{\frac{\kx}{2}}\sin{\frac{\ky}{2}}[\sx\rhoz\sigmax],
    \label{eq:on_plane_1}
\end{align}
and the next-nearest-neighbor hopping
\begin{align}
    H^{(2)}_{\parallel}(\bm{k})\Def
     & t_2
    (\cos{\kx}+\cos{\ky})[s_0\rho_0\sigma_0] \notag                                           \\
     & +v_{\mathrm{s}2}\Big(\sin{\kx}[\sx\rhoz\sigmaz]+\sin{\ky}[\sy\rhoz\sigmaz]\Big) \notag \\
     & +v'_{\mathrm{s}2}\Big(\sin{\kx}[\sy\rhoz\sigma_0]-\sin{\ky}[\sx\rhoz\sigma_0]\Big).
    \label{eq:on_plane_2}
\end{align}
For out-of-plane directions, we include the nearest-neighbor hopping
\begin{equation}
    H^{(1)}_{\perp}(\bm{k})\Def u_1\cos{\frac{\kz}{2}}[s_0\rhox\sigma_0]+u_2\sin{\frac{\kz}{2}}[s_0\rhoy\sigma_0],
    \label{eq:out_plane_1}
\end{equation}
the next-nearest-neighbor hopping
\begin{align}
    H^{(2)}_{\perp}(\bm{k})\Def
     & w_1\cos{\frac{\kx}{2}}\cos{\frac{\ky}{2}}\cos{\frac{\kz}{2}}[s_0\rhox\sigmax] \notag  \\
     & +w_2\cos{\frac{\kx}{2}}\cos{\frac{\ky}{2}}\cos{\frac{\kz}{2}}[\sz\rhox\sigmay] \notag \\
     & +w_3\cos{\frac{\kx}{2}}\cos{\frac{\ky}{2}}\sin{\frac{\kz}{2}}[s_0\rhoy\sigmax] \notag \\
     & +w_4\cos{\frac{\kx}{2}}\cos{\frac{\ky}{2}}\sin{\frac{\kz}{2}}[\sz\rhoy\sigmay] \notag \\
     & +w_5\cos{\frac{\kx}{2}}\sin{\frac{\ky}{2}}\sin{\frac{\kz}{2}}[\sy\rhox\sigmay] \notag \\
     & +w_5\sin{\frac{\kx}{2}}\cos{\frac{\ky}{2}}\sin{\frac{\kz}{2}}[\sx\rhox\sigmay] \notag \\
     & +w_6\sin{\frac{\kx}{2}}\cos{\frac{\ky}{2}}\cos{\frac{\kz}{2}}[\sx\rhoy\sigmay] \notag \\
     & +w_6\cos{\frac{\kx}{2}}\sin{\frac{\ky}{2}}\cos{\frac{\kz}{2}}[\sy\rhoy\sigmay],
    \label{eq:out_plane_2}
\end{align}
and the next-next-nearest-neighbor hopping
\begin{align}
    H^{(3)}_{\perp}(\bm{k})\Def
     & v_1
    (\cos{\kx}+\cos{\ky})\cos{\frac{\kz}{2}}[s_0\rhox\sigma_0] \notag \\
     & +v_2
    (\cos{\kx}+\cos{\ky})\sin{\frac{\kz}{2}}[s_0\rhoy\sigma_0],
    \label{eq:out_plane_3}
\end{align}
in addition to the intra-sublattice hopping
\begin{equation}
    H^{(4)}_{\perp}(\bm{k})\Def p_1\cos{\kz}[s_0\rho_0\sigma_0]+p_2\sin{\kz}[\sz\rhoz\sigmaz].
    \label{eq:out_plane_4}
\end{equation}
Here, we take the basis as $(c_{X,\up},c_{X,\down})\ (X=\mathrm{A,B,C,D})$.
The tight-binding Hamiltonian is then given by
\begin{multline}
    H_\wp(\bm{k})\Def H^{(1)}_{\parallel}(\bm{k})+H^{(2)}_{\parallel}(\bm{k}) \\
    +H^{(1)}_{\perp}(\bm{k})+H^{(2)}_{\perp}(\bm{k})+H^{(3)}_{\perp}(\bm{k})+H^{(4)}_{\perp}(\bm{k}).
    \label{eq:wp}
\end{multline}
Figure~\ref{fig:wp_ene} shows the surface energy spectrum of a semi-infinite system with a $(001)$ surface in the C--D layer.
The surface spectrum is computed using the recursive Green's function method~\cite{umerski1997closed}.
We confirm the characteristic wallpaper fermion surface states, and thus adopt $H_\wp(\bm{k})$ as a normal Hamiltonian throughout this work~\footnote{
    The terms $H^{(1)}_{\parallel}(\bm{k})$, $H^{(2)}_{\parallel}(\bm{k})$, $H^{(1)}_{\perp}(\bm{k})$, and $H^{(3)}_{\perp}(\bm{k})$ were derived in Ref.~\onlinecite{wieder2018wallpaper}.
}.

\subsection{Classification of pair potentials} \label{sec:pair_potential}

To describe the superconducting states, we define the BdG Hamiltonian as
\begin{equation}
    H_\scwp(\bm{k})\Def\qty[H_\wp(\bm{k})-\mu]\tauz+\hat{\Delta}_i\taux,
    \label{eq:BdG}
\end{equation}
where $\mu$ is the chemical potential and $\hat{\Delta}_i$ denotes the pair potential.
Here, $\tau_\nu$ ($\nu=0,\mathrm{x,y,z}$) are Pauli matrices in the Nambu (particle-hole) space.
We take the Nambu basis as $(c_{X,\up},c_{X,\down},c^\dagger_{X,\down},-c^\dagger_{X,\up})$.

\begin{table}
    \begin{ruledtabular}
        \centering
        \caption{
            Characters of the point group $\mathrm{D_{4h}}$ and the corresponding on-site pair potentials.
            We impose Fermi--Dirac statistics: $\sy\hat{\Delta}^\mathrm{t}_i\sy=\hat{\Delta}_i$.
            $\Delta_0$ is a momentum-independent constant.
        }
        \begin{tabular}{crrrcl}
            Irrep             & $2C_4$ & $2C'_2$ & $I$  & Basis                    & Pair potential                                \\ \hline
            $\mathrm{A_{1g}}$ & $1$    & $1$     & $1$  & $\kx^2+\ky^2,\ \kz^2$    & $\hat{\Delta}_1\Def\Delta_0s_0\rho_0\sigma_0$ \\
            $\mathrm{A_{2g}}$ & $1$    & $-1$    & $1$  & $\kx\ky(\kx^2-\ky^2)$    & $\hat{\Delta}_2\Def\Delta_0s_0\rho_0\sigmaz$  \\
            $\mathrm{A_{1u}}$ & $1$    & $1$     & $-1$ & $\kx\ky\kz(\kx^2-\ky^2)$ & $\hat{\Delta}_3\Def\Delta_0s_0\rhoz\sigmaz$   \\
            $\mathrm{A_{2u}}$ & $1$    & $-1$    & $-1$ & $\kz$                    & $\hat{\Delta}_4\Def\Delta_0s_0\rhoz\sigma_0$  \\
        \end{tabular}
        \label{tab:Delta}
    \end{ruledtabular}
\end{table}
Pair potentials are decomposed into irreducible representations (irreps) of the point group $\mathrm{D_{4h}}$, which is the rotational part of space group $P4/mbm$.
Restricting on-site pairings in this work, and imposing Fermi--Dirac statistics: $\sy\hat{\Delta}^\mathrm{t}_i\sy=\hat{\Delta}_i$, we identify four symmetry-distinct on-site pair potentials, $\hat{\Delta}_1$--$\hat{\Delta}_4$, listed in Table~\ref{tab:Delta}.

\section{Surface States} \label{sec:surface_state}
In this section, we analyze the $(001)$ surface states for each pairing symmetry based on symmetry-based and numerical approaches.

\subsection{Wallpaper fermions in superconducting states} \label{sec:wp_sc_state}

\begin{table}
    \begin{ruledtabular}
        \centering
        \caption{
            Compatibility relations between the irreps of the bulk point group $\mathrm{D_{4h}}$ and surface point group $\mathrm{C_{4v}}$.
            The rightmost column summarizes the superconducting gap structure in surface wallpaper fermions reported in Ref.~\onlinecite{yoda2026superconducting}.
        }
        \begin{tabular}{ccc}
            Irrep of $\mathrm{D_{4h}}$          & Irrep of $\mathrm{C_{4v}}$      & Gap structure                                                \\ \hline
            $\mathrm{A_{1g}}\ (\hat{\Delta}_1)$ & \multirow{2}{*}{$\mathrm{A}_1$} & \multirow{2}{*}{Fully gapped}                                \\
            $\mathrm{A_{2u}}\ (\hat{\Delta}_4)$ &                                 &                                                              \\ \hline
            $\mathrm{A_{2g}}\ (\hat{\Delta}_2)$ & \multirow{2}{*}{$\mathrm{A}_2$} & \multirow{2}{*}{Point nodes along the $\bar{\Gamma}\M$ line} \\
            $\mathrm{A_{1u}}\ (\hat{\Delta}_3)$ &                                 &                                                              \\
        \end{tabular}
        \label{tab:comprel}
    \end{ruledtabular}
\end{table}
In this work, by hybridization we mean the spectral reconstruction in which the branches derived from wallpaper fermions that remain gapless in the superconducting state away from the $\M$ point are continuously connected to the branches emanating from the Majorana Kramers pairs at the $\M$ point, giving rise to an unconventional surface spectrum with multiple zero-energy crossings.
To realize such hybridization, the wallpaper fermions need to remain gapless in the superconducting state.
Within the weak-coupling assumption, whether a given pairing opens a gap on the surface dispersion is determined solely by surface symmetry, i.e., the compatibility of irreps of energy bands in the normal state and pair potential in the superconducting state.
As shown in Ref.~\onlinecite{yoda2026superconducting}, we previously reported the superconducting gap structures in wallpaper fermions based on the $p4g$ surface symmetries.
Pair potentials belonging to the $\mathrm{A}_2$ representation of the point group $\mathrm{C_{4v}}$, the rotational part of the wallpaper group $p4g$, exhibit symmetry-enforced point nodes along the $\bar{\Gamma}\M$ lines, whereas those belonging to the $\mathrm{A}_1$ representation are fully gapped.
Using compatibility relations between the bulk point group $\mathrm{D_{4h}}$ and the surface point group $\mathrm{C_{4v}}$ in Table~\ref{tab:comprel}, we therefore conclude that the $\Delta_2$ and $\Delta_3$ pairings do not open a superconducting gap in the wallpaper fermion dispersion along the $\bar{\Gamma}\M$ lines.

\subsection{Majorana Kramers pairs in wallpaper fermion systems} \label{sec:majorana}

Next, we discuss the emergence of Majorana Kramers pairs for the $\Delta_2$ and $\Delta_3$ pairings at the $\M$ point.
Here, Majorana Kramers pairs refer to zero-energy boundary states that appear at the time-reversal-invariant momentum in the superconducting state as topological boundary modes of the BdG Hamiltonian.
They appear at the $\M$ point when the BdG Hamiltonian along the MA line, $(\kx,\ky)=(\pi,\pi),-\pi\leq\kz\leq\pi$ [Fig.~\ref{fig:model_BZ}\subref{fig:BZ}], is topologically nontrivial.
The BdG Hamiltonian~\eqref{eq:BdG} possesses a chiral symmetry: $\Gamma H_\scwp(\bm{k})\Gamma^\dag=-H_\scwp(\bm{k})$ ($\Gamma\Def\tauy$) and thus one may define a winding number~\cite{sato2011topology}.
However, we consider a time-reversal-invariant one-dimensional path, on which the conventional winding number is forced to vanish due to time-reversal symmetry~\cite{xiong2017anisotropica}.
We therefore exploit order-two crystalline symmetries that leave the MA line invariant~\cite{yamazaki2021magnetic}.

According to Ref.~\onlinecite{yamazaki2021magnetic}, the emergence of Majorana Kramers pairs is governed by one-dimensional topological invariants, which are determined by the space group symmetry, the $\bm{k}$ point, and the superconducting pairing symmetry.
The MA line is invariant under twofold rotation about the $z$ axis, $C_{2\mathrm{z}}\Def\{2_{001}|\bm{0}\}$, whose representation satisfies $D_{\bm{k}}(C_{2\mathrm{z}})^2=-1$.
Moreover, the pair potentials $\hat{\Delta}_2$ and $\hat{\Delta}_3$ are even under $C_{2\mathrm{z}}$.
Following the criterion derived in Refs.~\onlinecite{xiong2017anisotropica} and \onlinecite{yamazaki2021magnetic}, we can then define a $\Z$ topological invariant, the magnetic winding number,
\begin{equation}
    W[C_{2\mathrm{z}}]\Def\frac{i}{4\pi}\int_{-\pi}^{\pi}
    \dd{\kz}
    \Tr\qty[\Gamma[C_{2\mathrm{z}}]H^{-1}_\scwp(\bm{k})\pdv{H_\scwp(\bm{k})}{\kz}],
    \label{eq:mag_w}
\end{equation}
where $\Gamma[C_{2\mathrm{z}}]\Def i(D_{\bm{k}}(C_{2\mathrm{z}})\tau_0)\Gamma$ is the magnetic chiral operator.
Here, $D_{\bm{k}}(C_{2\mathrm{z}})\tau_0$ denotes the representation of $C_{2\mathrm{z}}$ for the BdG Hamiltonian~\eqref{eq:BdG}.

We now evaluate Eq.~\eqref{eq:mag_w} for the BdG Hamiltonian~\eqref{eq:BdG} with $\Delta_2$ and $\Delta_3$ pairing.
First of all, along the MA line in the space group $P4/mbm$, the magnetic winding number is restricted to $4\Z=\{0,\pm4,\pm8,\ldots\}$ because of four-dimensional irreps at the $\M$ point in the wallpaper group $p4g$~\cite{Aroyo2011-cr, Aroyo2006-bi1, Aroyo2006-bi2, elcoro2021magnetic, xu2020high}.
In addition, a set of necessary conditions for the magnetic winding number to take a nonzero value, formulated in terms of additional order-two crystalline symmetries, has been derived in Ref.~\onlinecite{xiong2017anisotropica}.
Based on these conditions, we find that $W[C_{2\mathrm{z}}]$ always vanishes for even-parity pairings under inversion \footnote{Reference~\onlinecite{xiong2017anisotropica} shows that all even-parity pairings have zero magnetic winding numbers.}, and the BdG Hamiltonian along the MA line is topologically trivial for $\Delta_2$ pairings.
In contrast, for $\Delta_3$ pairing, $W[C_{2\mathrm{z}}]$ can be nonzero.
Hence, combining the results of Sec.~\ref{sec:wp_sc_state} and the above analysis, we conclude that only for the $\Delta_3$ pairing, the gapless wallpaper fermions and the Majorana Kramers pairs can coexist along the $\bar{\Gamma}\M$ line.

\subsection{Hybridized state} \label{sec:hybridization}

In this subsection, we analyze the superconducting surface states for the $\Delta_3$ pairing.
The parameters of the tight-binding Hamiltonian~\eqref{eq:wp} are the same as those written in Fig.~\ref{fig:wp_ene}.
We set the chemical potential to $\mu=-0.75$ so that the Fermi level crosses the bulk spectra at the $\M$ point [Fig.~\ref{fig:wp_ene}], and fix the pairing amplitude to $\Delta_0=0.02$, assuming weak-coupling pairing.
For this parameter set, we obtain $W[C_{2\mathrm{z}}]=4$, and hence double Majorana Kramers pairs appear at the $\M$ point.

\begin{figure}
    \centering
    \includegraphics[keepaspectratio,scale=0.35]{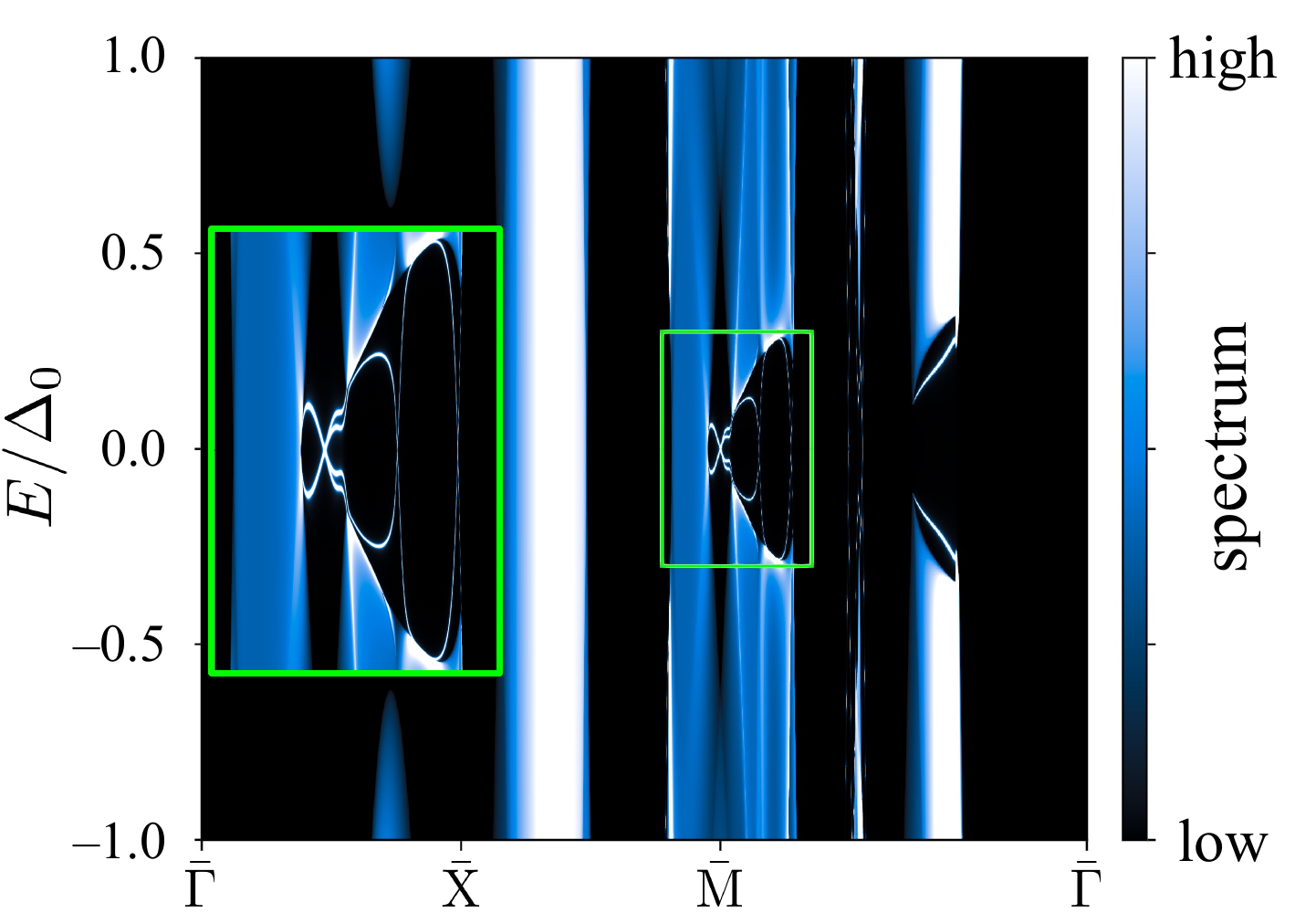}
    \label{fig:wpsc_ene}
    \caption{(Color online)
            Superconducting surface energy spectrum of the $\Delta_3$ pairing for a semi-infinite system with the $(001)$ surface on the C--D layers.
            The inset shows an enlarged view near the $\M$ point.
    }
    \label{fig:wpsc_ene}
\end{figure}
\begin{figure*}
    \centering
    \includegraphics[keepaspectratio,scale=0.5]{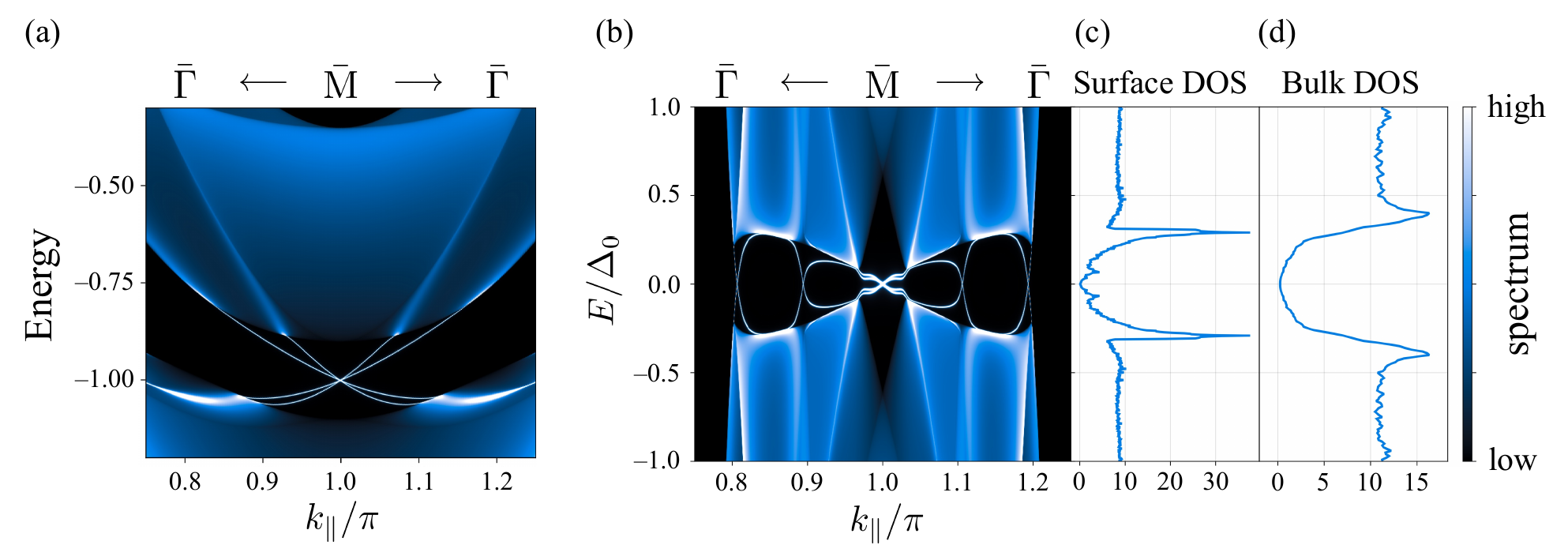}
    \caption{(Color online)
        Surface spectral function for a $(001)$ surface on the C--D layer along the $\bar{\Gamma}\M\bar{\Gamma}$ line in (a) the normal state and (b) the superconducting state with $\Delta_3$ pairing ($\Delta_0=0.02,\mu=-0.75$).
        (c) Surface and (d) bulk density of states for the $\Delta_3$ pairing.
    }
    \label{fig:result}
\end{figure*}
The superconducting surface spectrum for the $\Delta_3$ pairing along the $\bar{\Gamma}\X\M\bar{\Gamma}$ path is shown in Fig.~\ref{fig:wpsc_ene}.
Along the $\X\M$ line, the projected bulk spectrum remains gapless because symmetry-protected nodes exist on the bulk $\ky=\pi$ plane.
Thus, among these momentum paths, we focus in the following on the $\bar{\Gamma}\M$ line, where the Majorana Kramers pairs coexist with gapless wallpaper fermion spectra.
The corresponding spectrum along the $\bar{\Gamma}\M\bar{\Gamma}$ line is shown in Fig.~\ref{fig:result}\hyperref[fig:result]{(b)}.
Hereafter, $k_\parallel$ denotes the momentum parameter along the $\bar{\Gamma}\M\bar{\Gamma}$ line, defined by $(\kx,\ky)=(k_\parallel,k_\parallel)$.
With this convention, $k_\parallel/\pi=1$ at the $\M$ point.
The wallpaper fermion branches that cross the Fermi level in the normal state remain gapless for $k_\parallel/\pi \sim$ 0.8 (1.2) and 0.9 (1.1) in the superconducting state, consistent with the symmetry-enforced point nodes along the $\bar{\Gamma}\M$ line.
In addition, double Majorana Kramers pairs emerge at the $\M$ point, as expected from the nontrivial magnetic winding number $W[C_{2\mathrm{z}}]=4$.
Thus, for the $\Delta_3$ pairing, two distinct kinds of gapless surface states coexist in the superconducting state.
A salient feature is the presence of two types of hybridization: (i) between wallpaper fermions themselves for 0.8 (1.1) $< k_\parallel/\pi <$ 0.9 (1.2) and (ii) between wallpaper fermions and Majorana Kramers pairs for $0.9 < k_\parallel/\pi < 1.1$.
As a consequence, the surface spectrum exhibits a characteristic \textit{double-twisted} structure along the $\bar{\Gamma}\M$ line.
Such a spectral reconstruction is absent in conventional time-reversal-invariant topological superconductivity, where the low-energy surface state is solely composed of Majorana Kramers pairs.
The key ingredient is the persistence of crystalline-symmetry-protected wallpaper fermions in the superconducting state, which provides additional gapless channels that can hybridize with Majorana Kramers pairs.
We note that the \textit{double-twisted} surface spectrum is not symmetry-enforced for arbitrary parameter sets.
Rather, it appears in parameter regimes where the symmetry-protected wallpaper-fermion-derived gapless modes coexist at low energy with Majorana Kramers pairs.
If the relevant wallpaper-fermion modes do not cross the Fermi level, or if the one-dimensional invariant on the MA line is trivial, the \textit{double-twisted} spectrum does not appear.
Appendix~\ref{app:surface} summarizes the numerical results and theoretical discussions for the superconducting surface states for the other pairings.

We further compute the surface DOS per unit area [Fig.~\ref{fig:result}\hyperref[fig:result]{(c)}] and compare it with the bulk DOS per unit volume [Fig.~\ref{fig:result}\hyperref[fig:result]{(d)}].
The \textit{double-twisted} dispersion yields nearly flat segments in the surface spectrum [Fig.~\ref{fig:result}\hyperref[fig:result]{(b)}].
The segments around $0.9<k_\parallel/\pi<1.1$ lead to weak enhancements in the surface DOS around $E/\Delta_0\simeq\pm0.066$, while the segments around $0.8\ (1.1)<k_\parallel/\pi<0.9\ (1.2)$ produce sharp peaks around $E/\Delta_0\simeq\pm0.294$.
In the same energy window of these sharp peaks, the bulk DOS is also enhanced, indicating that the bulk states contribute to the spectral weight.
Nevertheless, the bulk DOS does not exhibit comparably sharp peaks.
This contrast suggests that the pronounced surface DOS peaks at $E/\Delta_0\simeq\pm0.294$ originate from combined surface and bulk contributions, and the surface component is comparable in magnitude to the bulk one.

For the present parameter set, the four peaks in the surface DOS can be traced back to the \textit{double-twisted} surface spectrum.
The double twist gives rise to two characteristic energy scales associated with regions of small group velocity, or van~Hove-like features, in the surface branches along the $\bar{\Gamma}\M$ lines.
At the same time, the appearance of these features as pronounced DOS peaks is not determined by symmetry alone.
Since the surface DOS is obtained by integrating the spectral weight over the full two-dimensional surface Brillouin zone, the peak positions, heights, and visibility depend on the detailed momentum dependence of the surface spectrum and on possible overlap with bulk states.

\section{Discussion} \label{sec:discussion}

In this section, we clarify how the superconducting surface state of topological nonsymmorphic crystalline insulators with wallpaper fermions discussed in Sec.~\ref{sec:surface_state} differs from those of a superconducting topological insulator $\mathrm{Cu}_x\mathrm{Bi}_2\mathrm{Se}_3$~\cite{hsieh2012Majorana} and a superconducting topological crystalline insulator $\mathrm{Sn}_{1-x}\mathrm{In}_x\mathrm{Te}$~\cite{hashimoto2015surface}.
Their surface spectra can be examined along a mirror-invariant line, where the energy eigenstates can be decomposed into mirror eigenspaces.
This enables a characterization in terms of mirror Chern number~\cite{teo2008surface, hsieh2012Majorana}, and allows us to point out what is unique to the wallpaper fermion case.

We briefly summarize the general properties of the mirror Chern number $n_{\mathcal{M}}$ based on Ref.~\onlinecite{hsieh2012Majorana}.
Let $n_{+i}$ and $n_{-i}$ denote the Chern numbers evaluated in the $+i$ and $-i$ mirror eigensectors, respectively.
The mirror Chern number is defined as $n_\mathcal{M}\Def(n_{+i}-n_{-i})/2$.
For a time-reversal-invariant system, time reversal exchanges the two mirror sectors, enforcing $n_{+i}=-n_{-i}$, and hence $n_\mathcal{M}=n_{+i}=-n_{-i}$.
The magnitude $|n_{\mathcal{M}}|$ gives the number of mirror-helical edge modes on a mirror-invariant boundary.
In addition, the sign $\eta\Def\sgn(n_\mathcal{M})=\sgn(n_{+i})=-\sgn(n_{-i})$ specifies the mirror helicity: in our convention, for $\eta>0$ ($\eta<0$), the $+i$ ($-i$) sector hosts right-going modes on the mirror-invariant line.

\begin{figure}
    \centering
    \includegraphics[keepaspectratio,scale=0.5]{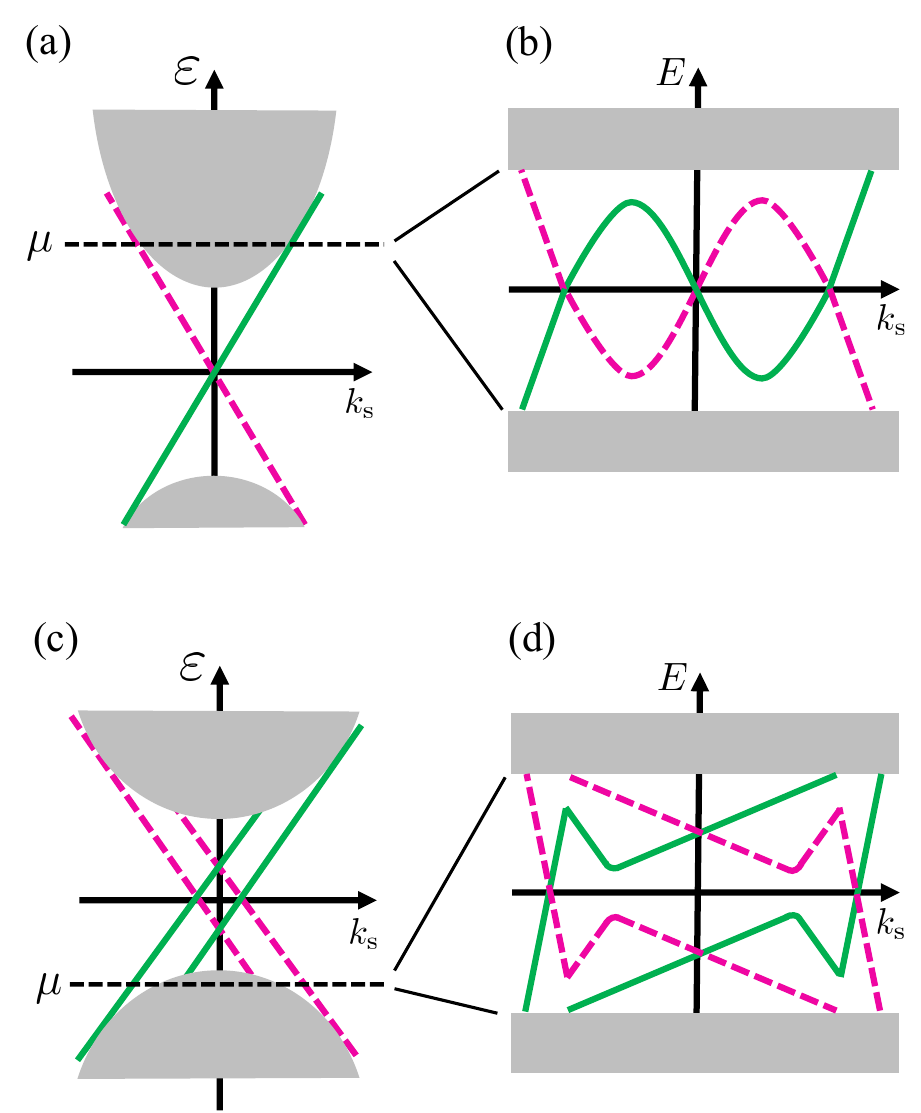}
    \caption{(Color online)
        Schematic surface dispersions for (a) a topological insulator, (b) its superconducting state, (c) a topological crystalline insulator, and (d) its superconducting state.
        Solid green (dotted magenta) lines indicate the dispersions belonging to $+i$ ($-i$) mirror eigenspace.
        $k_\mathrm{s}$ denotes the surface momentum.
    }
    \label{fig:STI_STCI}
\end{figure}
Figures~\ref{fig:STI_STCI}\hyperref[fig:STI_STCI]{(a)} and \ref{fig:STI_STCI}\hyperref[fig:STI_STCI]{(b)} schematically illustrate the surface states of a topological insulator and its superconducting state \cite{hsieh2012Majorana}, while Figs.~\ref{fig:STI_STCI}\hyperref[fig:STI_STCI]{(c)} and \ref{fig:STI_STCI}\hyperref[fig:STI_STCI]{(d)} illustrate those of a topological crystalline insulator and its superconducting state \cite{hashimoto2015surface}.
For the superconducting topological insulators [Fig.~\ref{fig:STI_STCI}\hyperref[fig:STI_STCI]{(b)}], the surface spectrum on the mirror line hosts a single helical pair with positive mirror helicity in our convention, corresponding to $n_{\mathcal{M}}=+1$.
Similarly, the superconducting topological crystalline insulators [Fig.~\ref{fig:STI_STCI}\hyperref[fig:STI_STCI]{(d)}] exhibit two helical pairs with positive mirror helicity, corresponding to $n_{\mathcal{M}}=+2$.
In both cases, the direction of propagation is locked to the mirror eigenvalue; the surface states are therefore characterized by a nonzero mirror Chern number and a well-defined mirror helicity.

\begin{figure}
    \centering
    \includegraphics[keepaspectratio,scale=0.35]{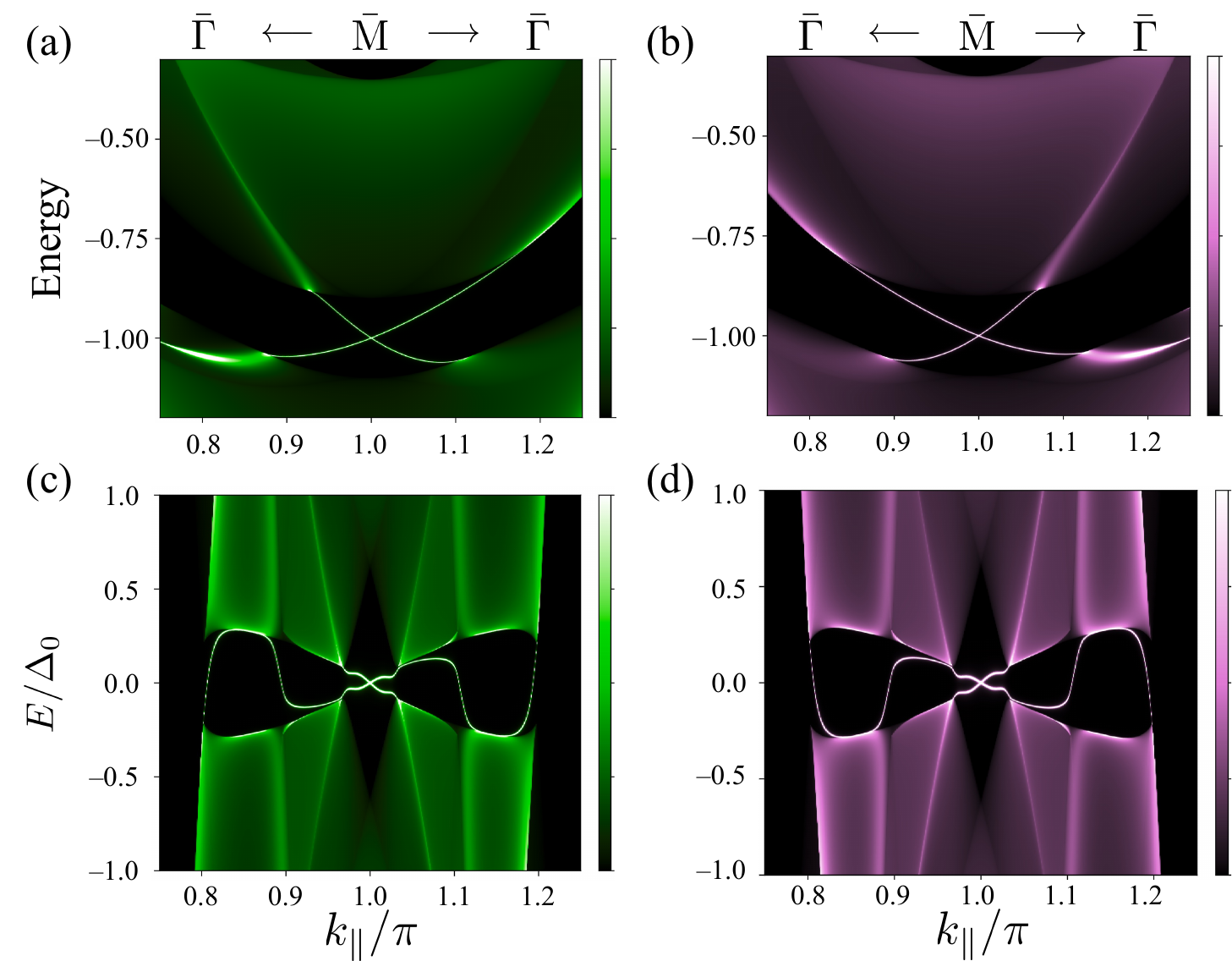}
    \caption{(Color online)
        Energy spectra along the $\bar{\Gamma}\M\bar{\Gamma}$ line, resolved into mirror eigensectors.
        Normal state spectra in the (a) $+i$ and (b) $-i$ sectors.
        Superconducting state spectra for the $\Delta_3$ pairing in the (c) $+i$ and (d) $-i$ sectors.
    }
    \label{fig:wp_mirror}
\end{figure}
In contrast, the superconducting wallpaper fermion behaves qualitatively differently.
In Sec.~\ref{sec:hybridization}, we focused on the $\bar{\Gamma}\M\bar{\Gamma}$ line, which is invariant under the mirror operation $\{m_{1\bar{1}0}|1/2\ 1/2\ 0\}$.
Accordingly, on this mirror-invariant line, $(\kx,\ky)=(k_\parallel,k_\parallel)$, the normal and BdG Hamiltonian can be block diagonalized into mirror eigenspaces as
    \begin{equation}
        H_\mathrm{\alpha}(k_\parallel,k_\parallel,\kz)\to\mqty(H^{+}_\mathrm{\alpha}(k_\parallel,k_\parallel,\kz) & O \\ O & H^{-}_\mathrm{\alpha}(k_\parallel,k_\parallel,\kz)),
    \end{equation}
    where $H^{\pm}_\alpha$ denotes the Hamiltonian projected onto the $\pm i$ eigenspace of the mirror operation $\{m_{1\bar{1}0}|1/2\ 1/2\ 0\}$, with $\alpha=\mathrm{wp,\ scwp}$ referring to the normal and BdG Hamiltonian, respectively.
Consequently, the normal state spectrum [Fig.~\ref{fig:result}\hyperref[fig:result]{(a)}] and the superconducting spectrum [Fig.~\ref{fig:result}\hyperref[fig:result]{(b)}] can be resolved into the $+i$ and $-i$ sectors as shown in Figs.~\ref{fig:wp_mirror}\hyperref[fig:wp_mirror]{(a)} and \ref{fig:wp_mirror}\hyperref[fig:wp_mirror]{(b)} and Figs.~\ref{fig:wp_mirror}\hyperref[fig:wp_mirror]{(c)} and \ref{fig:wp_mirror}\hyperref[fig:wp_mirror]{(d)}, respectively.
From Figs.~\ref{fig:wp_mirror}\hyperref[fig:wp_mirror]{(c)} and \ref{fig:wp_mirror}\hyperref[fig:wp_mirror]{(d)}, each mirror eigensector contains both right- and left-going branches along $\bar{\Gamma}\M\bar{\Gamma}$ lines.
Therefore, the propagation direction is not locked to the mirror eigenvalue, which implies that the net Chern number within each sector cancels, resulting in $n_\mathcal{M}=n_{+i}=n_{-i}=0$.
Equivalently, the surface spectrum is mirror-helicity-free in the sense that the mirror eigenvalue does not uniquely determine the direction of motion.

This mirror-helicity-free character can be traced back to the mirror-resolved band structure in the normal state.
As shown in Figs.~\ref{fig:wp_mirror}\hyperref[fig:wp_mirror]{(a)} and \ref{fig:wp_mirror}\hyperref[fig:wp_mirror]{(b)}, each mirror eigensector already contains counter-propagating branches in the normal state, so that the mirror Chern number vanishes.
In the weak-coupling limit, the superconducting state inherits this property~\cite{hsieh2012Majorana}, and the BdG mirror Chern number therefore remains zero.
Thus, the absence of mirror helicity in the present system is not a direct consequence of particle-hole symmetry alone, but originates from the vanishing mirror Chern number of the normal-state wallpaper-fermion bands.

Although the absence of mirror helicity itself is not a directly observable quantity, it may be reflected indirectly in response phenomena of a magnetically gapped surface.
To make this point more concrete, let us consider a magnetic-proximity setup in which a ferromagnetic insulator is attached to the same $(001)$ surface analyzed above.
If the exchange field induced by the ferromagnetic insulator opens a full gap in the surface spectrum, the resulting gapped surface may exhibit a thermal Hall response.
For superconducting topological insulators such as $\mathrm{Cu}_x\mathrm{Bi}_2\mathrm{Se}_3$, an analogous ferromagnetic-insulator junction has been shown to yield a quantum thermal Hall response whose magnitude is quantized in terms of the mirror Chern number as \cite{shimizu2015quantum}
\begin{equation}
    |\kappa_{\mathrm{xy}}|=\frac{\pi^2 k_{\mathrm{B}}^2T}{6h}|n_{\mathcal{M}}|.
\end{equation}
In the present wallpaper-fermion system, however, $n_{\mathcal{M}}=0$.
Therefore, if the magnetically gapped surface response were governed solely by the mirror Chern number, one would expect the vanishing thermal Hall contribution associated with mirror helicity.
Clarifying whether a ferromagnetic-insulator junction can fully gap the present wallpaper-fermion surface, and whether the resulting thermal Hall response is controlled by other topological indices or by details of the induced surface mass terms, requires a separate analysis and is left for future work.

Finally, we briefly comment on a possible stabilization mechanism for the $\Delta_3$ pairing discussed above.
Since the $\hat{\Delta}_3=\Delta_0 s_0\rhoz\sigmaz$ pair potential changes sign between the in-plane nearest-neighbor sublattices [A(C) and B(D)] and also between the out-of-plane nearest-neighbor sublattices [A(B) and C(D)], repulsive electron interactions on corresponding bonds may favor this pairing.
A definitive conclusion, however, requires a microscopic analysis, such as solving the gap equation.

\section{Conclusion} \label{sec:conclusion}
In this work, we constructed a symmetry-consistent tight-binding model and investigated the $(001)$ surface states in superconducting topological nonsymmorphic crystalline insulators with wallpaper fermions.
Based on a symmetry-based theoretical analysis, we classified four types of on-site pairings allowed by $\mathrm{D_{4h}}$ symmetry, and found that the $\Delta_3$ pairing supports both gapless wallpaper fermions and double Majorana Kramers pairs along the $\bar{\Gamma}\M$ lines.
In this pairing, hybridization between these quasiparticles yields a characteristic \textit{double-twisted} surface spectrum.
In addition, the surface DOS exhibits peak structures, reflecting the interplay between symmetry-protected wallpaper fermions and Majorana Kramers pairs.
We further showed that the resulting surface spectrum is mirror-helicity-free, in sharp contrast to a superconducting topological insulator $\mathrm{Cu}_x\mathrm{Bi}_2\mathrm{Se}_3$ and a superconducting topological crystalline insulator $\mathrm{Sn}_{1-x}\mathrm{In}_x\mathrm{Te}$, which are characterized by a nonzero mirror Chern number.
These findings help to elucidate and explore the nontrivial nature of topological superconductivity in topological materials.

\acknowledgments
This work is supported by JSPS KAKENHI for Grants (Grants Nos. JP24H00853 and JP25K07224).

\section*{Data Availability}
The data that support the findings of this article are publicly available~\cite{data}.

\appendix

\section{Representation matrices} \label{app:rep}
In this section, we briefly describe the derivation of the representation matrices $D_{\bm{k}}(g)$ in Eqs.~\eqref{eq:four-rot}--\eqref{eq:inversion}.
We use the Fourier transformation of the real-space fermion operators,
\begin{equation}
    c^\dag_{\bm{k},s,X}\Def\frac{1}{\sqrt{N}}\sum_{\bm{n}}e^{i\bm{k}\cdot(\bm{n}+\bm{\tau}_X)}c^\dag_{s,\bm{n}+\bm{\tau}_X},
\end{equation}
where $s=\up,\down,\ X=\mathrm{A,B,C,D}$.
Here, $\bm{n}$ denotes a lattice vector, $N$ is the number of unit cells, and $\bm{\tau}_X$ specifies the position of sublattice $X$ from $\bm{n}$.
In this convention, for a space group operation $g=\{p_g|\bm{a}_g\}$, the representation matrix can be decomposed into spin, sublattice (SL), and phase factor parts as~\cite{satow2025symmetry}
\begin{multline}
    [D_{\bm{k}}(g)]_{s'X',sX} \\
    =[D^{(\mathrm{spin})}(p_g)]_{s',s}[D^{(\mathrm{SL})}(g)]_{X',X}e^{-i(p_g\bm{k})\cdot\bm{a}_g}.
\end{multline}
Here, $D^{(\mathrm{spin})}(p_g)$ describes the action on the spin space rotation and is determined by the point group part $p_g$, while $D^{(\mathrm{SL})}(g)$ is the permutation matrix representing the sublattice transformation.

For the generators considered in the main text, the spin part is given by
\begin{align}
     & D^{(\mathrm{spin})}(4^+_{001})=\exp(-i\frac{\pi}{4}\sz)=\frac{1}{\sqrt{2}}(s_0-i\sz), \\
     & D^{(\mathrm{spin})}(2_{100})=\exp(-i\frac{\pi}{2}\sx)=-i\sx,                          \\
     & D^{(\mathrm{spin})}(I)=s_0.
\end{align}
Using the sublattice basis $(\mathrm{A,B,C,D})$, the permutation matrix is derived as
\begin{align}
     & D^{(\mathrm{SL})}(\{4^+_{001}|\bm{0}\})=\rho_0\sigma_0,  \\
     & D^{(\mathrm{SL})}(\{2_{100}|1/2\ 1/2\ 0\})=\rhox\sigmax, \\
     & D^{(\mathrm{SL})}(\{I|\bm{0}\})=\rhox\sigma_0.
\end{align}
Finally, by evaluating the phase factor $e^{-i(p_g\bm{k})\cdot\bm{a}_g}$ for each generator, we obtain the explicit forms of $D_{\bm{k}}(g)$ given in Eqs.~\eqref{eq:four-rot}--\eqref{eq:inversion}.

\section{Surface states for the other pair potentials} \label{app:surface}
In this section, we summarize the superconducting surface states for four symmetry-distinct pair potentials.

\subsection{Comparison with surface states for each pair potential} \label{app:comparison}
\begin{figure}
    \centering
    \includegraphics[keepaspectratio,scale=0.5]{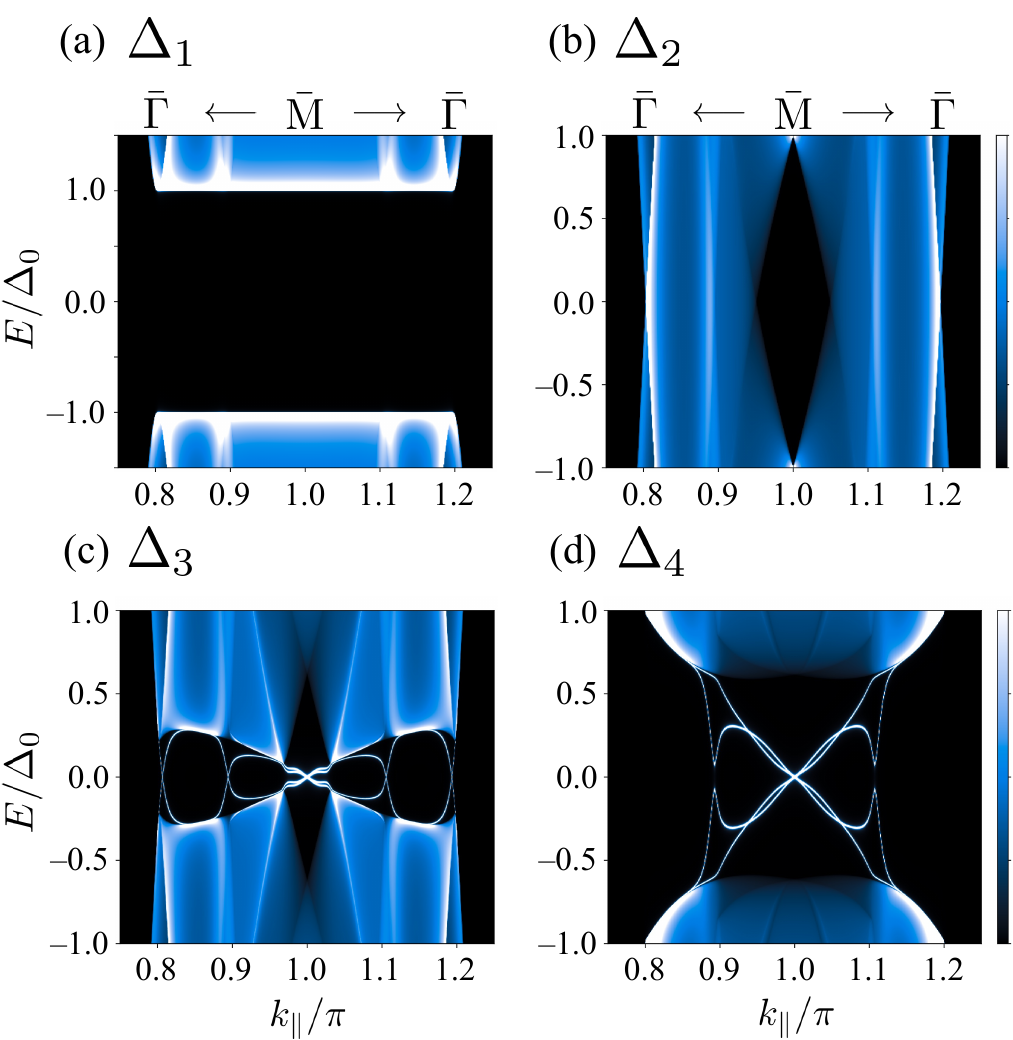}
    \caption{(Color online)
        Superconducting surface states along the $\bar{\Gamma}\M\bar{\Gamma}$ line for (a) $\Delta_1$, (b) $\Delta_2$, (c) $\Delta_3$, and (d) $\Delta_4$ pairings.
        The parameters are the same as those used in Sec.~\ref{sec:hybridization}.
    }
    \label{fig:all}
\end{figure}
As shown in Fig.~\ref{fig:all}\hyperref[fig:wp_mirror]{(a)}, for the $\Delta_1$ pairing, which corresponds to $s$-wave superconductivity, both bulk states and wallpaper fermions are fully gapped.
For the $\Delta_2$ pairing [Fig.~\ref{fig:all}\hyperref[fig:wp_mirror]{(b)}], symmetry-protected line nodes exist on the diagonal plane ($\kx=\ky$) in the bulk Brillouin zone, and therefore no superconducting gap opens in the projected bulk spectrum along the $\bar{\Gamma}\M$ direction.
In addition, we confirm that Majorana Kramers pairs protected by $W[C_{2\mathrm{z}}]$ do not appear at the $\M$ point.
In contrast, for the $\Delta_4$ pairing [Fig.~\ref{fig:all}\hyperref[fig:wp_mirror]{(d)}], a superconducting gap opens in both the bulk and surface states, and Majorana Kramers pairs emerge at the $\M$ point (see Appendix~\ref{app:Delta4}).
From these results, we find that hybridization between double Majorana Kramers pairs and wallpaper fermions occurs only for the $\Delta_3$ pairing.

\subsection{Topological nature for the \texorpdfstring{$\Delta_4$}{Delta4} pairing} \label{app:Delta4}
In this subsection, we discuss that the $\Delta_4$ pairing has the $\mathbb Z_2$ topological invariant, applying the criterion derived in Ref.~\onlinecite{yamazaki2021magnetic}.
We consider the glide with respect to the $yz$ plane
\begin{align}
    M_\mathrm{x}\Def\{m_{100}|1/2\ 1/2\ 0\},
\end{align}
which is an order-two symmetry that leaves the MA ($k_x=k_y=\pi$) line invariant.
The above operation is squared to be $M_{\mathrm x}^2 = \{ {}^d\!E | 0\ 1\ 0 \}$,
where ${}^d\!E$ denotes the ``$2\pi$ rotation'' in the double group, hence
\begin{align}
    D_{\bm{k}}(M_\mathrm{x}^2) = D_{M_\mathrm{x}\bm{k}}(M_\mathrm{x}) D_{\bm{k}}(M_\mathrm{x}) = 1,
\end{align}
on the MA line.
Choosing the gauge as $H_\wp(\boldsymbol k) = H_\wp(\boldsymbol k + \boldsymbol G)$ with $\boldsymbol G$ a reciprocal lattice vector, $D_{M_\mathrm{x} \boldsymbol k}(M_\mathrm{x}) = D_{\boldsymbol k}(M_\mathrm{x})$ on the MA line, the above relation reduces to
\begin{align}
    D_{\boldsymbol k}(M_\mathrm{x})^2 = 1.
\end{align}
The pair potential is even parity for the glide,
\begin{align}
    D_{\bm{k}}(M_\mathrm{x})\hat{\Delta}_4D_{\bm{k}}(M_\mathrm{x})^\dag=+\hat{\Delta}_4.
\end{align}
The above two relations correspond to the type D (nonsymmorphic) in Table II in Ref.~\onlinecite{yamazaki2021magnetic}.

The other surface symmetry operation $M_\mathrm{y}\Def\{m_{010}|1/2\ 1/2\ 0\}$ anticommutes with $M_\mathrm{x}$, $\{M_\mathrm{x},M_\mathrm{y}\}=0$.
The $\Delta_4$ pairing is odd parity for the inversion.
Applying the Fermi-surface criterion summarized in Fig.~5 of Ref.~\onlinecite{yamazaki2021magnetic}, one finds that a $\Z_2$ topological invariant $\nu_{\pm}[M_\mathrm{x}]$ can be defined in each eigenspace of $M_\mathrm{x}$ and in the weak-coupling limit, these invariants are given by
\begin{equation}
    \nu_{\pm}[M_\mathrm{x}]=\frac{N_\occ(\pi)-N_\occ(0)}{4} \pmod 2,
\end{equation}
where $N_\occ(\kz)$ denotes the number of occupied bands at momentum $\kz$ in the normal state.

In our case, the bulk bands at the M and A point are necessarily fourfold degenerate, owing to four-dimensional irreps in the space group $P4/mbm$~\cite{Aroyo2011-cr, Aroyo2006-bi1, Aroyo2006-bi2, elcoro2021magnetic, xu2020high}.
Therefore, if a Fermi surface exists along the MA line, we obtain $\nu_+[M_\mathrm{x}]=\nu_-[M_\mathrm{x}]=1$, implying a Majorana Kramers pair in each eigenspace.
Hence, for the $\Delta_4$ pairing, double Majorana Kramers pairs appear at the $\M$ point, as shown in Fig.~\ref{fig:all}\hyperref[fig:wp_mirror]{(d)}.
The above discussion also applies upon exchanging $M_{\mathrm x}$ and $M_{\mathrm y}$. Therefore, we can similarly define $\nu_\pm[M_\mathrm{y}]$.
However, it is not independent of $\nu_\pm[M_{\mathrm x}]$ owing to $[M_{\mathrm x}, M_{\mathrm y}] \ne 0$.

\bibliography{ref}

@article{hao2011surface,
  title = {Surface Spectral Function in the Superconducting State of a Topological Insulator},
  author = {Hao, Lei and Lee, T. K.},
  year = {2011},
  month = apr,
  journal = {Phys. Rev. B},
  volume = {83},
  numpages = {13},
  pages = {134516},
  publisher = {American Physical Society},
  doi = {10.1103/PhysRevB.83.134516},
  urldate = {2024-09-23}
}

@article{hashimoto2015surface,
  title = {Surface Electronic State of Superconducting Topological Crystalline Insulator},
  author = {Hashimoto, Tatsuki and Yada, Keiji and Sato, Masatoshi and Tanaka, Yukio},
  year = {2015},
  month = nov,
  journal = {Phys. Rev. B},
  volume = {92},
  numpages = {17},
  pages = {174527},
  publisher = {American Physical Society},
  doi = {10.1103/PhysRevB.92.174527},
  urldate = {2024-09-22}
}

@article{hsieh2012Majorana,
  title = {Majorana {Fermions} and {Exotic} {Surface} {Andreev} {Bound} {States} in {Topological} {Superconductors}: {Application} to {Cu}${}_x${Bi}\textsubscript{2}{Se}\textsubscript{3}},
  author = {Hsieh, Timothy H. and Fu, Liang},
  journal = {Phys. Rev. Lett.},
  volume = {108},
  issue = {10},
  pages = {107005},
  numpages = {5},
  year = {2012},
  month = {Mar},
  publisher = {American Physical Society},
  doi = {10.1103/PhysRevLett.108.107005}
}

@article{kawakami2018topological,
  title = {Topological {{Crystalline Materials}} of ${J}=3/2$ {{Electrons}}: {{Antiperovskites}}, {{Dirac Points}}, and {{High Winding Topological Superconductivity}}},
  shorttitle = {Topological {{Crystalline Materials}} of {{J}} = 3 / 2 {{Electrons}}},
  author = {Kawakami, Takuto and Okamura, Tetsuya and Kobayashi, Shingo and Sato, Masatoshi},
  year = {2018},
  month = nov,
  journal = {Phys. Rev. X},
  volume = {8},
  numpages = {4},
  pages = {041026},
  issn = {2160-3308},
  doi = {10.1103/PhysRevX.8.041026},
  urldate = {2024-09-13},
  langid = {english}
}

@article{lu2015crossed,
  title = {Crossed {{Surface Flat Bands}} of {{Weyl Semimetal Superconductors}}},
  author = {Lu, Bo and Yada, Keiji and Sato, Masatoshi and Tanaka, Yukio},
  year = {2015},
  month = mar,
  journal = {Phys. Rev. Lett.},
  volume = {114},
  numpages = {9},
  pages = {096804},
  issn = {0031-9007, 1079-7114},
  doi = {10.1103/PhysRevLett.114.096804},
  urldate = {2025-08-19},
  copyright = {http://link.aps.org/licenses/aps-default-license},
  langid = {english}
}

@article{mizuno2023hall,
  title = {Hall Effect of Ferro/Antiferromagnetic Wallpaper Fermions},
  author = {Mizuno, Koki and Yamakage, Ai},
  year = {2023},
  month = jun,
  journal = {Phys. Rev. B},
  volume = {107},
  numpages = {23},
  pages = {235301},
  publisher = {American Physical Society},
  doi = {10.1103/PhysRevB.107.235301},
  urldate = {2024-09-23}
}

@article{novak2013unusual,
  title = {Unusual Nature of Fully Gapped Superconductivity in {{In-doped SnTe}}},
  author = {Novak, Mario and Sasaki, Satoshi and Kriener, Markus and Segawa, Kouji and Ando, Yoichi},
  year = {2013},
  month = oct,
  journal = {Phys. Rev. B},
  volume = {88},
  numpages = {14},
  pages = {140502(R)},
  doi = {10.1103/PhysRevB.88.140502}
}

@article{shimizu2015quantum,
  title = {Quantum Thermal {{Hall}} Effect of {{Majorana}} Fermions on the Surface of Superconducting Topological Insulators},
  author = {Shimizu, Yosuke and Yamakage, Ai and Nomura, Kentaro},
  year = {2015},
  month = may,
  journal = {Phys. Rev. B},
  volume = {91},
  numpages = {19},
  pages = {195139},
  publisher = {American Physical Society},
  doi = {10.1103/PhysRevB.91.195139},
  urldate = {2024-12-21}
}

@article{wieder2018wallpaper,
  title = {Wallpaper Fermions and the Nonsymmorphic {{Dirac}} Insulator},
  author = {Wieder, Benjamin J. and Bradlyn, Barry and Wang, Zhijun and Cano, Jennifer and Kim, Youngkuk and Kim, Hyeong-Seok D. and Rappe, Andrew M. and Kane, C. L. and Bernevig, B. Andrei},
  year = {2018},
  month = jul,
  journal = {Science},
  volume = {361},
  number = {6399},
  pages = {246--251},
  publisher = {American Association for the Advancement of Science},
  doi = {10.1126/science.aan2802},
  urldate = {2025-06-02}
}

@article{yamakage2012theory,
  title = {Theory of Tunneling Conductance and Surface-State Transition in Superconducting Topological Insulators},
  author = {Yamakage, Ai and Yada, Keiji and Sato, Masatoshi and Tanaka, Yukio},
  year = {2012},
  month = may,
  journal = {Phys. Rev. B},
  volume = {85},
  numpages = {18},
  pages = {180509(R)},
  issn = {1098-0121, 1550-235X},
  doi = {10.1103/PhysRevB.85.180509},
  urldate = {2024-09-13},
  copyright = {http://link.aps.org/licenses/aps-default-license},
  langid = {english}
}

@article{yamakage2013theory,
  title = {Theory of Tunneling Spectroscopy in a Superconducting Topological Insulator},
  author = {Yamakage, Ai and Yada, Keiji and Sato, Masatoshi and Tanaka, Yukio},
  year = {2013},
  month = nov,
  journal = {Physica C: Superconductivity},
  volume = {494},
  pages = {20--23},
  issn = {09214534},
  doi = {10.1016/j.physc.2013.04.019},
  urldate = {2024-10-02},
  langid = {english}
}

@article{yamazaki2021magnetic,
  title = {Magnetic Response of {{Majorana Kramers}} Pairs with an Order-Two Symmetry},
  author = {Yamazaki, Yuki and Kobayashi, Shingo and Yamakage, Ai},
  year = {2021},
  month = mar,
  journal = {Phys. Rev. B},
  volume = {103},
  numpages = {9},
  pages = {094508},
  publisher = {American Physical Society},
  doi = {10.1103/PhysRevB.103.094508},
}

@article{zhou2021glide,
  title = {Glide Symmetry Protected Higher-Order Topological Insulators from Semimetals with Butterfly-like Nodal Lines},
  author = {Zhou, Xiaoting and Hsu, Chuang-Han and Huang, Cheng-Yi and Iraola, Mikel and Ma{\~n}es, Juan L. and Vergniory, Maia G. and Lin, Hsin and Kioussis, Nicholas},
  year = {2021},
  month = dec,
  journal = {npj Comput. Mater.},
  volume = {7},
  number = {1},
  pages = {202},
  publisher = {Nature Publishing Group},
  issn = {2057-3960},
  doi = {10.1038/s41524-021-00672-9},
  urldate = {2025-08-19},
  copyright = {2021 The Author(s)},
  langid = {english}
}

@article{hwang2023magnetic,
  title = {Magnetic Wallpaper {{Dirac}} Fermions and Topological Magnetic {{Dirac}} Insulators},
  author = {Hwang, Yoonseok and Qian, Yuting and Kang, Junha and Lee, Jehyun and Ryu, Dongchoon and Choi, Hong Chul and Yang, Bohm-Jung},
  year = {2023},
  month = apr,
  journal = {npj Comput. Mater.},
  volume = {9},
  number = {1},
  pages = {65},
  issn = {2057-3960},
  doi = {10.1038/s41524-023-01018-3}
}

@article{ryu2020wallpaper,
  title = {Wallpaper {{Dirac Fermion}} in a {{Nonsymmorphic Topological Kondo Insulator}}: {Pu}{B}${}_4$},
  author = {Ryu, Dong-Choon and Kim, Junwon and Choi, Hongchul and Min, Byung Il},
  year = {2020},
  month = nov,
  journal = {J. Am. Chem. Soc.},
  volume = {142},
  number = {45},
  pages = {19278--19282},
  publisher = {American Chemical Society},
  issn = {0002-7863},
  doi = {10.1021/jacs.0c09442}
}

@article{mizuno2025magnon,
  title = {Magnon-mediated superconductivity at the interface between a ferromagnetic insulator and a topological crystalline insulator with wallpaper fermions},
  author = {Mizuno, Koki and Yamakage, Ai},
  journal = {Phys. Rev. B},
  volume = {112},
  issue = {3},
  pages = {035303},
  numpages = {11},
  year = {2025},
  month = {Jul},
  publisher = {American Physical Society},
  doi = {10.1103/sknv-r7wq}
}

@article{Aroyo2011-cr,
	author = {Aroyo, M. I. and Perez-Mato, J. M. and Orobengoa, D. and Tasci, E. and De La Flor, G. and Kirov, A.},
	title = {Crystallography online: Bilbao crystallographic server},
	year = {2011},
	journal = {Bulgarian Chemical Communications},
	volume = {43},
	number = {2},
	pages = {183--197}
}

@article{Aroyo2006-bi1,
  title       = {Bilbao {Crystallographic} {Server}: I. {Databases} and crystallographic computing programs},
  author      = {Mois Ilia Aroyo and Juan Manuel Perez-Mato and Cesar Capillas and Eli Kroumova and Svetoslav Ivantchev and Gotzon Madariaga and Asen Kirov and Hans Wondratschek},
  pages       = {15--27},
  volume      = {221},
  number      = {1},
  journal     = {Zeitschrift f\"{u}r Kristallographie - Crystalline Materials},
  doi         = {doi:10.1524/zkri.2006.221.1.15},
  year        = {2006}
}

@article{Aroyo2006-bi2,
  author   = {Aroyo, Mois I. and Kirov, Asen and Capillas, Cesar and Perez-Mato, J. M. and Wondratschek, Hans},
  title    = {{Bilbao Crystallographic Server. II. Representations of crystallographic point groups and space groups}},
  journal  = {Acta Cryst. A},
  year     = {2006},
  volume   = {62},
  number   = {2},
  pages    = {115--128},
  month    = {Mar},
  doi      = {10.1107/S0108767305040286}
}

@article{elcoro2021magnetic,
  title={Magnetic topological quantum chemistry},
  author={Elcoro, Luis and Wieder, Benjamin J and Song, Zhida and Xu, Yuanfeng and Bradlyn, Barry and Bernevig, B Andrei},
  journal={Nat. Commun.},
  volume={12},
  number={1},
  pages={5965},
  year={2021},
  doi= {10.1038/s41467-021-26241-8},
  publisher={Nature Publishing Group UK London}
}

@article{xu2020high,
  title={High-throughput calculations of magnetic topological materials},
  author={Xu, Yuanfeng and Elcoro, Luis and Song, Zhi-Da and Wieder, Benjamin J and Vergniory, Maia G and Regnault, Nicolas and Chen, Yulin and Felser, Claudia and Bernevig, B Andrei},
  journal={Nature},
  volume={586},
  number={7831},
  pages={702--707},
  year={2020},
  doi={10.1038/s41586-020-2837-0},
  publisher={Nature Publishing Group UK London}
}

@article{Aroyo2014brillouin,
  author = "Aroyo, Mois I. and Orobengoa, Danel and de la Flor, Gemma and Tasci, Emre S. and Perez-Mato, J. Manuel and Wondratschek, Hans",
  title = "{Brillouin-zone database on the Bilbao Crystallographic Server}",
  journal = "Acta Cryst. A",
  year = "2014",
  volume = "70",
  number = "2",
  pages = "126--137",
  month = "Mar",
  doi = {10.1107/S205327331303091X},
}

@article{yonezawa2018nematic,
  title={Nematic {Superconductivity} in {Doped} {Bi}\textsubscript{2}{Se}\textsubscript{3} {Topological} {Superconductors}},
  author={Yonezawa, Shingo},
  journal={Condens. Matter},
  volume={4},
  number={1},
  pages={2},
  year={2018},
  publisher={MDPI},
  doi={10.3390/condmat4010002}
}

@article{sasaki2012odd,
  title = {Odd-{Parity} {Pairing} and {Topological} {Superconductivity} in a {Strongly} {Spin-Orbit} {Coupled} {Semiconductor}},
  author = {Sasaki, Satoshi and Ren, Zhi and Taskin, A. A. and Segawa, Kouji and Fu, Liang and Ando, Yoichi},
  journal = {Phys. Rev. Lett.},
  volume = {109},
  issue = {21},
  pages = {217004},
  numpages = {5},
  year = {2012},
  month = {Nov},
  publisher = {American Physical Society},
  doi = {10.1103/PhysRevLett.109.217004}
}

@article{wang2016observation,
  title={Observation of superconductivity induced by a point contact on 3{D} {Dirac} semimetal {Cd}\textsubscript{3}{As}\textsubscript{2} crystals},
  author={Wang, He and Wang, Huichao and Liu, Haiwen and Lu, Hong and Yang, Wuhao and Jia, Shuang and Liu, Xiong-Jun and Xie, XC and Wei, Jian and Wang, Jian},
  journal={Nat. Mater.},
  volume={15},
  number={1},
  pages={38--42},
  year={2016},
  publisher={Nature Publishing Group UK London},
  doi={10.1038/nmat4456}
}

@article{aggarwal2016unconventional,
  title={Unconventional superconductivity at mesoscopic point contacts on the 3{D} {Dirac} semimetal {Cd}\textsubscript{3}{As}\textsubscript{2}},
  author={Aggarwal, Leena and Gaurav, Abhishek and Thakur, Gohil S and Haque, Zeba and Ganguli, Ashok K and Sheet, Goutam},
  journal={Nat. Mater.},
  volume={15},
  number={1},
  pages={32--37},
  year={2016},
  publisher={Nature Publishing Group UK London},
  doi={10.1038/nmat4455}
}

@article{wang2017discovery,
  title={Discovery of tip induced unconventional superconductivity on {Weyl} semimetal},
  author={Wang, He and Wang, Huichao and Chen, Yuqin and Luo, Jiawei and Yuan, Zhujun and Liu, Jun and Wang, Yong and Jia, Shuang and Liu, Xiong-Jun and Wei, Jian and Wang, Jian},
  journal={Sci. Bulletin},
  volume={62},
  number={6},
  pages={425--430},
  year={2017},
  publisher={Elsevier},
  doi={10.1016/j.scib.2017.02.009}
}

@article{he2013full,
  title = {Full superconducting gap in the doped topological crystalline insulator {Sn}${}_{0.6}${In}${}_{0.4}${Te}},
  author = {He, L. P. and Zhang, Z. and Pan, J. and Hong, X. C. and Zhou, S. Y. and Li, S. Y.},
  journal = {Phys. Rev. B},
  volume = {88},
  issue = {1},
  pages = {014523},
  numpages = {4},
  year = {2013},
  month = {Jul},
  publisher = {American Physical Society},
  doi = {10.1103/PhysRevB.88.014523}
}

@article{kobayashi2015topological,
  title = {Topological {Superconductivity} in {Dirac} {Semimetals}},
  author = {Kobayashi, Shingo and Sato, Masatoshi},
  journal = {Phys. Rev. Lett.},
  volume = {115},
  issue = {18},
  pages = {187001},
  numpages = {5},
  year = {2015},
  month = {Oct},
  publisher = {American Physical Society},
  doi = {10.1103/PhysRevLett.115.187001}
}

@article{umerski1997closed,
  title = {Closed-form solutions to surface {Green's} functions},
  author = {Umerski, A.},
  journal = {Phys. Rev. B},
  volume = {55},
  issue = {8},
  pages = {5266--5275},
  numpages = {0},
  year = {1997},
  month = {Feb},
  publisher = {American Physical Society},
  doi = {10.1103/PhysRevB.55.5266}
}

@article{xiong2017anisotropica,
  title = {Anisotropic {{Magnetic Responses}} of {{Topological Crystalline Superconductors}}},
  author = {Xiong, Yuansen and Yamakage, Ai and Kobayashi, Shingo and Sato, Masatoshi and Tanaka, Yukio},
  year = {2017},
  month = feb,
  journal = {Crystals},
  volume = {7},
  number = {2},
  pages = {58},
  publisher = {Multidisciplinary Digital Publishing Institute},
  issn = {2073-4352},
  doi = {10.3390/cryst7020058},
  }

@article{mao2022third,
  title={Third-order topological insulators with wallpaper fermions in {Tl}\textsubscript{4}{Pb}{Te}\textsubscript{3} and {Tl}\textsubscript{4}{Sn}{Te}\textsubscript{3}},
  author={Mao, Ning and Wang, Hao and Dai, Ying and Huang, Baibiao and Niu, Chengwang},
  journal={npj Comput. Mater.},
  volume={8},
  number={1},
  pages={154},
  year={2022},
  doi = {10.1038/s41524-022-00839-y}
}

@article{teo2008surface,
  title = {Surface states and topological invariants in three-dimensional topological insulators: {Application} to {Bi}${}_{1-x}${Sb}${}_x$},
  author = {Teo, Jeffrey C. Y. and Fu, Liang and Kane, C. L.},
  journal = {Phys. Rev. B},
  volume = {78},
  issue = {4},
  pages = {045426},
  numpages = {15},
  year = {2008},
  month = {Jul},
  publisher = {American Physical Society},
  doi = {10.1103/PhysRevB.78.045426}
}

@article{qi2011topological,
  title = {Topological insulators and superconductors},
  author = {Qi, Xiao-Liang and Zhang, Shou-Cheng},
  journal = {Rev. Mod. Phys.},
  volume = {83},
  issue = {4},
  pages = {1057--1110},
  numpages = {0},
  year = {2011},
  month = {Oct},
  publisher = {American Physical Society},
  doi = {10.1103/RevModPhys.83.1057},
}

@article{ivanov2001non-abelian,
  title = {Non-{Abelian} {Statistics} of {Half}-{Quantum} {Vortices} in $p$-{Wave} {Superconductors}},
  author = {Ivanov, D. A.},
  journal = {Phys. Rev. Lett.},
  volume = {86},
  issue = {2},
  pages = {268--271},
  numpages = {0},
  year = {2001},
  month = {Jan},
  publisher = {American Physical Society},
  doi = {10.1103/PhysRevLett.86.268},
}

@article{kitaev2006anyons,
  title = {Anyons in an exactly solved model and beyond},
  author = {Kitaev, Alexei},
  journal = {Ann. Phys.},
  volume = {321},
  issue = {1},
  pages = {2--111},
  year = {2006},
  doi = {10.1016/j.aop.2005.10.005},
}

@article{nayak2008non-abelian,
  title = {Non-{Abelian} anyons and topological quantum computation},
  author = {Nayak, Chetan and Simon, Steven H. and Stern, Ady and Freedman, Michael and Das Sarma, Sankar},
  journal = {Rev. Mod. Phys.},
  volume = {80},
  issue = {3},
  pages = {1083--1159},
  numpages = {0},
  year = {2008},
  month = {Sep},
  publisher = {American Physical Society},
  doi = {10.1103/RevModPhys.80.1083},
}

@article{sato2011topology,
  title = {Topology of {Andreev} bound states with flat dispersion},
  author = {Sato, Masatoshi and Tanaka, Yukio and Yada, Keiji and Yokoyama, Takehito},
  journal = {Phys. Rev. B},
  volume = {83},
  issue = {22},
  pages = {224511},
  numpages = {22},
  year = {2011},
  month = {Jun},
  publisher = {American Physical Society},
  doi = {10.1103/PhysRevB.83.224511},
}

@article{yoda2026superconducting,
  title = {Superconducting {Gap} {Structures} in {Wallpaper} {Fermion} {Systems}},
  author = {Yoda, Kaito and Yamakage, Ai},
  journal = {J. Low Temp. Phys.},
  volume = {222},
  issue = {2},
  pages = {44},
  year = {2026},
  month = {Feb},
  doi = {10.1007/s10909-026-03368-w},
}

@article{sato-ando2017topological,
doi = {10.1088/1361-6633/aa6ac7},
year = {2017},
volume = {80},
number = {7},
pages = {076501},
author = {Sato, Masatoshi and Ando, Yoichi},
title = {Topological superconductors: a review},
journal = {Rep. Prog. Phys.},
}

@article{satow2025symmetry,
  title = {Symmetry-adapted models for multifold fermions with spin-orbit coupling},
  author = {Satow, Koki and Yamakage, Ai},
  journal = {Phys. Rev. B},
  volume = {112},
  issue = {19},
  pages = {195206},
  numpages = {13},
  year = {2025},
  month = {Nov},
  publisher = {American Physical Society},
  doi = {10.1103/m8c6-yvt3},
}

@article{chiu2016classification,
  title = {Classification of topological quantum matter with symmetries},
  author = {Chiu, Ching-Kai and Teo, Jeffrey C. Y. and Schnyder, Andreas P. and Ryu, Shinsei},
  journal = {Rev. Mod. Phys.},
  volume = {88},
  issue = {3},
  pages = {035005},
  numpages = {63},
  year = {2016},
  month = {Aug},
  publisher = {American Physical Society},
  doi = {10.1103/RevModPhys.88.035005},
}

@article{beenakker2013search,
   author = "Beenakker, C.W.J.",
   title = "Search for {Majorana} {Fermions} in {Superconductors}", 
   journal= "Annu. Rev. Condens. Matter Phys.",
   year = "2013",
   volume = "4",
   number = "Volume 4, 2013",
   pages = "113-136",
   doi = "10.1146/annurev-conmatphys-030212-184337",
   publisher = "Annual Reviews",
  }

@article{sato2016majorana,
author = {Sato ,Masatoshi and Fujimoto ,Satoshi},
title = {Majorana {Fermions} and {Topology} in {Superconductors}},
journal = {J. Phys. Soc. Jpn.},
volume = {85},
number = {7},
pages = {072001},
year = {2016},
doi = {10.7566/JPSJ.85.072001}
}

@article{alicea2012new,
doi = {10.1088/0034-4885/75/7/076501},
year = {2012},
month = {jun},
publisher = {IOP Publishing},
volume = {75},
number = {7},
pages = {076501},
author = {Alicea, Jason},
title = {New directions in the pursuit of {Majorana} fermions in solid state systems},
journal = {Rep. Prog. Phys.},
}

@article{aguado2017majorana,
  title={Majorana quasiparticles in condensed matter},
  author={Aguado, Ram{\'o}n},
  journal={Riv. Nuovo Cim.},
  volume={40},
  number={11},
  pages={523--593},
  year={2017},
  publisher={Springer},
  doi={10.1393/ncr/i2017-10141-9}
}

@article{mizushima2016symmetry,
author = {Mizushima ,Takeshi and Tsutsumi ,Yasumasa and Kawakami ,Takuto and Sato ,Masatoshi and Ichioka ,Masanori and Machida ,Kazushige},
title = {Symmetry-{Protected} {Topological} {Superfluids} and {Superconductors} ---{From} the {Basics} to ${}^3${He}---},
journal = {J. Phys. Soc. Jpn.},
volume = {85},
number = {2},
pages = {022001},
year = {2016},
doi = {10.7566/JPSJ.85.022001},
}

@misc{data,
author = {Yoda, Kaito and Yamakage, Ai},
title = {Data for ``{Double}-twisted surface spectrum from hybridized {Majorana} {Kramers} pairs and wallpaper fermions''},
year = {2026},
howpublished = {NAGOYA Repository},
note = {Available at \url{https://nagoya.repo.nii.ac.jp}}
}

\end{document}